\documentclass[11pt]{article}
\usepackage{cite,epic,eepic,euscript,verbatim,amsmath,amsthm,amssymb,amsfonts,afterpage,float,bm,stmaryrd,paralist,cancel,color,fancyhdr,graphics,graphicx,epsf,hyperref,makeidx,epsfig}
\usepackage{multirow}
\usepackage{subfigure,epstopdf}
\usepackage{array}
\numberwithin{equation}{section}
\newcommand{\reff}[1]{{\rm (\ref{#1})}}

\newcommand{\R}{\mathbb{R}}            % real numbers
\newcommand{\ve}{\varepsilon}          % varepsilon
\newcommand{\be}{\mathbf e}            % boldface e
\newcommand{\bx}{\mathbf x}            % boldface x
\def\XXint#1#2#3{{\setbox0=\hbox{$#1{#2#3}{\int}$}
\vcenter{\hbox{$#2#3$}}\kern-.51\wd0}}

\newcommand{\td}{\tilde}

\begin{document}
\title{Hysteresis and Linear Stability Analysis on Multiple Steady-State Solutions to the Poisson--Nernst--Planck equations with Steric Interactions}
%: Numerical Approaches
\author{Jie Ding\thanks{
Department of Mathematics and Mathematical Center for Interdiscipline Research, Soochow University, 1 Shizi Street, Suzhou 215006, Jiangsu, China. Email: 20174007003@stu.suda.edu.cn.}
\and
Hui Sun\thanks{
Department of Mathematics and Statistics, California State University, Long Beach, CA, U. S. A. Email: Paul.Sun@csulb.edu.}
\and
Shenggao Zhou\thanks{
Corresponding author. Department of Mathematics and Mathematical Center for Interdiscipline Research, Soochow University, 1 Shizi Street, Suzhou 215006, Jiangsu, China. Email: sgzhou@suda.edu.cn.
}
}
%%Submission uses .tex file and .eps figures. Otherwise, cannot generate figures.
\maketitle
\begin{abstract}
In this work, we numerically study linear stability of multiple steady-state solutions to a type of steric Poisson--Nernst--Planck (PNP) equations with Dirichlet boundary conditions, which are applicable to ion channels. With numerically found multiple steady-state solutions,  we obtain $S$-shaped current-voltage and current-concentration curves, showing hysteretic response of ion conductance to voltages and boundary concentrations with memory effects. Boundary value problems are proposed to locate bifurcation points and predict the local bifurcation diagram near bifurcation points on the $S$-shaped curves.  Numerical approaches for linear stability analysis are developed to understand the stability of the steady-state solutions that are only numerically available. Finite difference schemes are proposed to solve a derived eigenvalue problem involving differential operators. The linear stability analysis reveals that the $S$-shaped curves have two linearly stable branches of different conductance levels and one linearly unstable intermediate branch, exhibiting classical bistable hysteresis. As predicted in the linear stability analysis, transition dynamics, from a steady-state solution on the unstable branch to a one on the stable branches, are led by perturbations associated to the mode of the dominant eigenvalue. Further numerical tests demonstrate that the finite difference schemes proposed in the linear stability analysis are second-order accurate. Numerical approaches developed in this work can be applied to study linear stability of a class of time-dependent problems around their steady-state solutions that are computed numerically.

\bigskip
%\noindent\textbf{AMS subject classifications:} 35K61, 35Q92, 37M20, 78A35\\

\noindent \textbf{Keywords:}
Steric Poisson--Nernst--Planck Equations; Multiple Solutions;  Linear Stability Analysis; Bistability; Current-Voltage Curve; Bifurcation Diagram
\end{abstract}

{\allowdisplaybreaks
%%%%%%%%%%%%%%%%%%%%%%%%%%%%%%%%%%%%%%%%%%%%%%%%%%%%%%%%%%%%%%%%%%%%%%%
\section{Introduction}\label{s:Introduction}
Ion channels play crucial roles in regulating various biological functions, such as exchanging ions across cell membranes and maintaining membrane excitability~\cite{IonChanel_HandbookCRC15}. Ion permeation through channels is sensitively related to the applied transmembrane voltages, ionic concentrations of cells and extracellular medium, etc.~\cite{Sigworth_QRB94, Cui_PflugersArch94}. Recent years have seen a growing interest in characterizing hysteretic response of ion permeation to applied voltages and concentrations~\cite{Bezrukov_JCP06, Das_PRE2012, Cuello_PNAS17}. For instance, ion conductance through voltage-gated channels follows different paths when applied voltages increase and decrease periodically~\cite{Andersson_MathBiosci2010,FologeaBBActa_2011,Das_PRE2012}. Such a hysteresis phenomenon takes place when the frequency of oscillating applied voltages is comparable to the relaxation timescale of transitions between conductance states~\cite{Bezrukov_JCP06,FologeaBBActa_2011,Das_PRE2012}.

As a continuum mean-field model, the Poisson--Nernst--Planck (PNP) theory has been widely used in the description of ion permeation through channels~\cite{BarChenBob_SIAP92, SN:SIAM:09}. In such a theory, the Poisson equation governs the electrostatic potential due to charges stemming both from mobile ions and fixed charges in the channel. The Nernst--Planck equations describe the diffusion and convection of ions in gradients of the electrostatic potential. The nonlinear coupling between the electrostatic potential and ionic concentrations poses a challenge in studying the problem analytically and numerically. Mathematical theories on the PNP equations have been developed in the literature of semiconductor physics~\cite{PMarkowich_Book}. Global existence results on solutions to steady-state PNP equations have been established for certain initial and boundary conditions~\cite{PMarkowich_Book}, whereas the uniqueness of the solution is mainly limited to the case of very small applied voltages. Numerical simulations have evidenced the existence of multiple steady-state solutions to the PNP equation in one dimension. For instance, multiple steady-state solutions are studied numerically and asymptotically in the case of the PNP equations with piecewise constant fixed charges of large magnitude~\cite{Ward_SIAP91}. Singular perturbation methods, such as matched asymptotic expansions~\cite{BarChenBob_SIAP97, BazantChuBayly_SIAP05, BazantSteric_PRE07, SingerGillBob_ESIAM08, WangHeHuang_PRE14, JLXZhou_SIAP18, BobLiu_SIADS08} and geometric singular perturbation theory~\cite{WLiu_SIAP05, EL07, WLiu_JDE09, JiLiuZhang_SIAP15, BobLiuXu_Nonlinearity15}, have been developed for the PNP equations.  Such asymptotic methods are often based on an assumption that a small singular parameter arises from the scale separation between the Debye length and channel dimension. For instance,  time-dependent matched asymptotic analysis is developed for the PNP equations to characterize the diffuse-charge dynamics in electrochemical systems~\cite{BazantDiffChg_PRE04, BazantSteric_PRE07}.

%The asymptotic analysis is further extended for a modified PNP theory to study steric effects on the charging dynamics in electrolyte cells~\cite{BazantSteric_PRE07}.

%More recently, the time-dependent matched asymptotic analysis is applied to investigate the profound impact of dielectric boundary effects on the complex interfacial structures in the vicinity of electrodes, using a modified PNP model with the dielectric self-energy correction~\cite{}.

Despite its success in wide applications, the PNP theory is derived based on the mean-field approximation that neglects steric effects and ion-ion correlations~\cite{BazantReview_ACIS09}. To address these issues,  more complicated models have been developed to account for ionic steric effects, ion-ion correlations, etc.~\cite{KBA:PRE:2007a, Li_SIMA09, Li_Nonlinearity09, MaXu_JCP14, ZhouWangLi_PRE11, LiLiuXuZhou_Nonliearity13, BZLu_BiophyJ11, BZLu_JSP16, LinBob_CMS14, HyonLiuBob_CMS10, XuShengLiu_CMS14}. For instance, the steric effect of ions can be taken into account through including the solvent entropy to the free energy of a charged system. Another approach is to incorporate ion-ion interactions described by the Lennard-Jones (LJ) integrals to the free energy~\cite{HyonLiuBob_CMS10, BobHyonLiu_JCP10}. To avoid computationally intractable integro-differential equations, local approximations of the nonlocal LJ integrals are proposed to obtain local PNP models with steric effects~\cite{LinBob_CMS14, HorngLinLiuBob_JPCB12, Gavish_PhysD18}. Modification added to the PNP theory in such a local model can also be understood as inclusion of hard-sphere interactions through second-order virial coefficients~\cite{ZhouJiangDoi_PRL17, DingWangZhou_JCP19}.  To validate the local steric PNP model, it is used to predict ion transport and selectivity of ionic channels~\cite{HorngLinLiuBob_JPCB12, GavishLiuEisenberg_JPCB18}.

The study on multiple steady-state solutions to the local steric PNP model has also attracted much attention in recent years~\cite{LinBob_Nonlinearity15,HungMihn_arXiv15, Gavish_PhysD18, DSWZhou_CICP18, GavishLiuEisenberg_JPCB18}. When zero-flux boundary conditions are considered, the steady-state Nernst-Planck type of equations can be reduced to a system of algebraic equations, which defines generalized Boltzmann distributions. Rigorous analyses prove that there are multiple steady-state solutions to such a reduced system coupled with the Poisson equation~\cite{LinBob_Nonlinearity15,HungMihn_arXiv15, Gavish_PhysD18}. It should be noted that the analysis for the zero-flux boundary conditions is not applicable to the scenario of ion permeation through channels. If multiple steady-state solutions present, a hysteresis loop can emerge in the current-voltage ($I$-$V$) characteristic curve of ion channels~\cite{DSWZhou_CICP18}.  It is desirable to perform stability analysis of the steady-state solutions in the sense of time evolution. Linear stability analysis is a powerful tool to understand the stability of solutions under perturbations~\cite{RosinSun_PRE13, Yochelis_JPCC14, Yochelis_PCCP14, LiSunZhou_SIAP15, GavishYochelis_JPCLett16, Gavish_PhysD18}. However, most of studies on linear stability analysis are carried out around steady states that have closed-form solutions. We develop general numerical approaches to perform linear stability analysis on solutions that are only numerically available, and apply them to numerically study the stability of steady-state solutions of the steric PNP equations in the context of ion channels.

In this work, we study linear stability of multiple steady-state solutions to a type of steric PNP equations that has been used to predict ion permeation through ion channels~\cite{LinBob_CMS14, HorngLinLiuBob_JPCB12}. The current-voltage ($I$-$V$) characteristic curve of ion channel is determined by finding the steady-state solutions of the steric PNP equations with Dirichlet boundary conditions for ionic concentrations and the electrostatic potential. The existence of multiple steady-state solutions gives rise to an $S$-shaped $I$-$V$ curve with a hysteresis loop, which demonstrates the dependence of ion conductance on varying applied voltages with memory effects. In addition, we find hysteretic response of ion conductance to boundary concentrations, indicating that ion conductance may not only depend on intracellular and extracellular ionic concentrations but also their history values.

We propose boundary value problems to predict the location of bifurcation points and corresponding local bifurcation diagram on the $S$-shaped $I$-$V$ curve.  Numerical approaches for linear stability analysis are developed to understand the stability of the steady-state solutions that are only numerically available. Finite difference schemes are proposed to discretize a derived eigenvalue problem involving differential operators. By solving the eigenvalue problem, the linear stability analysis reveals that both the lower and upper branches of the $S$-shaped $I$-$V$ curve are linearly stable and the intermediate branch is linearly unstable, exhibiting a classical bistable hysteresis diagram. Furthermore, we numerically study the transition dynamics in nonlinear regime, from a steady-state solution on the unstable branch to a one on the stable branches. As predicted by the linear stability analysis, such transition dynamics is led by perturbations associated to the mode of the dominant eigenvalue. Numerical simulations also demonstrate that the local bifurcation diagram predicted by the second derivative of $V$ with respect to $I$ agrees with the $I$-$V$ curve. In addition, we perform numerical tests to show the convergence of the finite difference schemes proposed in the linear stability analysis.

The rest of the paper is organized as follows. In section~\ref{s:ModelDescrip}, we describe our governing equations.  In section~\ref{s:SteadyStates}, we present our methods to solve for steady-state solutions and bifurcation points.  In section~\ref{s:LinStbAn}, we perform linear stability analysis on steady-state solutions. Section~\ref{s:NumRes} is devoted to the presentation of our numerical results. Finally, we draw conclusions in section~\ref{s:Conclusions}.

\section{Model}\label{s:ModelDescrip}
%Consider a charged system occupying a bounded, open, connected domain $\Omega$ with a smooth boundary $\partial\Omega$.

Consider a charged system occupying a bounded domain $\Omega$ with a smooth boundary $\partial\Omega$. Dirichlet boundary conditions are considered on $\partial \Omega$ with boundary data given by $V_0$. The electrostatic potential $\psi$ satisfies the following boundary-value problem of the Poisson equation:
\begin{equation*}
\left\{
\begin{aligned}
- \nabla\cdot\varepsilon_0 \varepsilon_r\nabla \psi &=  \sum_{l=1} ^M z_l ec_l + \rho_f ~\mbox{ in }\Omega,\\
\psi&=V_0 \qquad\qquad~~~\mbox{ on } \partial\Omega,
\end{aligned}
\right.
\end{equation*}
where $c_l$ is the ion concentration for the $l$-th species, $z_l$ is the valence, $M$ is the number of ion species, $e$ is the elementary charge, $\ve_0$ is the vacuum permittivity, $\ve_r$ is the relative dielectric constant, and $\rho_f$ is the fixed charge density. The free-energy functional of the charged system is given by~\cite{LiuQiaoLu_SIAP18, LinBob_CMS14, HorngLinLiuBob_JPCB12}:
\begin{equation}\label{F}
\begin{aligned}
F[c_1, c_2, \cdots, c_M]=&\frac{1}{2}\int_{\Omega}\left(\sum_{l=1}^M z_l e c_l+\rho_f\right)\psi d\bx + \beta^{-1} \int_{\Omega}  \sum_{l=1}^M c_l \left[\log\left(\Lambda^3 c_l\right)-1\right] d\bx   \\
&\qquad \qquad+\sum_{l=1}^M \sum_{r=1}^M  \frac{\beta^{-1}}{2}\int\int_{\Omega} c_l \omega_{lr} c_r d\bx-\frac{1}{2}\int_{\partial\Omega}\varepsilon_0\varepsilon_r\frac{\partial\psi}{\partial\mathbf{n}}V_0dS,\\
\end{aligned}
\end{equation}
where $\beta$ is the inverse thermal energy, $\Lambda$ is the thermal de Broglie wavelength, and $\omega_{ij}$ are the steric interaction coefficients, depending on the size of $i$th and $j$th ionic species. The coefficients can be obtained by  introducing spatially band-limited functions to truncate the (spatial) frequency range of the singular Lennard-Jones interactions between ions~\cite{LinBob_CMS14, HorngLinLiuBob_JPCB12}, and are given by 
$$\omega_{ij}=\varepsilon_{ij}(a_i+a_j)^{12}S_{\delta},$$
where $\varepsilon_{ij}$ is the energy coupling constant between the $i$th and $j$th ionic species, $a_i$ and $a_j$ are the corresponding radii of ions, and $S_{\delta}= \frac{\omega_d}{12-d}\delta^{-12+d}$ with $\omega_d$ being the surface area of $d$ dimensional unit ball and $\delta$ being a small number for cutoff length.
%In addition, the interaction coefficient matrix
%\begin{equation*}\label{stericM}
%W=(\omega_{ij})_{M\times M}
%\end{equation*}
%is a symmetric matrix with diagonal elements representing self steric interactions of ions of the same species and off-diagonal elements representing cross steric interactions between ions of different species. 
Note that the steric interaction coefficients $\omega_{ij}$ can be alternatively defined via the second-order virial coefficients for hard spheres~\cite{ZhouJiangDoi_PRL17}.

The evolution of ion concentrations is described by the continuity equation
\[
\partial_t c_l = -\nabla\cdot J_l,
\]
where the flux $J_l$ is given by
$$J_l=-\beta D_l c_l \nabla \left(\frac{\delta F}{\delta c_l}\right),~l=1,2,\cdots,M.$$
Here $D_l$ is the diffusion constant of $l$th ionic species. This gives rise to the Nernst--Planck equations
$$\partial_t c_l  = D_l\nabla\cdot\left(\nabla c_l+z_le\beta c_l\nabla\psi+ c_l\sum_{r=1}^M \omega_{lr} \nabla c_r\right),~ l=1, 2, \cdots, M.$$
We introduce a macroscopic length scale $L$, a characteristic concentration $c_0$, a characteristic screening length $\lambda_{\rm D}= \sqrt{\frac{\ve_0 \ve_r}{2\beta e^2 c_0}}$, and the following dimensionless variables~\cite{BazantDiffChg_PRE04, BazantSteric_PRE07}
\begin{equation*}\label{Rescale}
\td \bx = \frac{\bx}{L}, ~ \td D_l =\frac{D_l}{D_0}, ~\td t =\frac{t D_0}{L\lambda_{\rm D}}, ~\td c_l =\frac{c_l}{c_0}, ~ \td \rho_f= \frac{\rho_f}{ec_0}, ~ \td \omega_{lr}=c_0\omega_{lr}, \td \psi=\beta e \psi.
\end{equation*}
After rescaling, we obtain by dropping all the tildes the nondimensionalized Poisson--Nernst--Planck equations with steric interactions
\begin{equation}\label{SPNP}
 \left\{
\begin{aligned}
&\partial_t c_l =\gamma_l \nabla \cdot \left( \nabla  c_l + z_lc_l \nabla \psi + c_l\sum_{r=1}^M\omega_{lr} \nabla c_r \right), ~ l=1,2,\cdots, M, \\
&-\Delta \psi =\kappa\left(\sum_{l=1}^Mz_lc_l + \rho_f\right),\\
\end{aligned}
\right.
\end{equation}
where $\gamma_l=\frac{\lambda_{\rm D}}{L}D_l$ and $\kappa=\frac{L^2}{2\lambda^2_{\rm D}}.$

We apply the steric PNP equations~\reff{SPNP} to study ion conductance through a perfect cylinder shaped ion channel. By assuming radial and azimuthal symmetry, we reduce~\reff{SPNP} into one-dimensional equations on a rescaled domain $\Omega=(-1,1)$:
\begin{equation}\label{SPNP1D}
\left\{
\begin{aligned}
&\partial_t c_l =\gamma_l \partial_x  \left ( \partial_x c_l + z_l c_l \partial_x \psi+ c_l\sum_{r=1}^Mw_{lr} \partial_x c_r \right) ,~ l=1, \dots, M,\\
&- \partial_{xx} \psi =\kappa \left( \sum_{l=1} ^M z_l c_l + \rho_f \right).\\
\end{aligned}
\right.
\end{equation}
Note that the dimension reduction used here is a special case of a more general one-dimensional reduction with a cross-sectional area factor, which allows more general geometry~\cite{EL07, WLiu_JDE09}.

We are interested in the case that two ends of an ion channel are connected to extracellular and intracellular ionic solution reservoirs.  To model an applied transmembrane potential and extracellular and intracellular bulk concentrations in reservoirs,  we employ the following boundary conditions~\cite{HorngLinLiuBob_JPCB12, GardnerJones_JTB11, WLiu_SIAP05, BobLiuXu_Nonlinearity15}
\begin{equation}\label{Bcs}
\begin{aligned}
&c_l(t, -1)=c_l^L ~\mbox{and}~ c_l(t, 1)=c_l^R, ~~  l=1, \dots, M,\\
&\psi(-1)=0 ~\mbox{and}~ \psi(1)=V,\\
\end{aligned}
\end{equation}
where $V$ is an applied potential difference across the channel and the bulk concentrations $c_l^L$ and $c_l^R$ at boundaries satisfy neutrality conditions
\begin{equation*}\label{NeuCon}
\sum_{l=1}^M z_l c_l^L = \sum_{l=1}^M z_l c_l^R = 0.
\end{equation*}
To describe fixed charges along a channel, we consider the following fixed charge distribution
\begin{equation}\label{rhof}
\rho_f(x)=\rho_i~~\mbox{for } x\in(x_{i-1},x_i),
\end{equation}
where $i=1,2,\cdots,K$, $x_0=-1$, $x_K=1$, and $\rho_1=\rho_K=0$. To quantify ion conductance through a channel, we define the net ionic current
\begin{equation}\label{Current}
I = \sum_{l=1}^M z_l J_l.
\end{equation}

We remark that ions often have strong correlations due to crowded confinement in ion channels. Although the steric interactions can be regarded as a part of ion correlations of short range,  the treatment on ion correlations in this work is incomplete.  As shown in~\cite{HorngLinLiuBob_JPCB12}, however, selective ion permeation through channels can be predicted to some extent by the inclusion of the pairwise interaction term $\sum_{l=1}^M \sum_{r=1}^M  \frac{\beta^{-1}}{2}\int\int_{\Omega} c_l \omega_{lr} c_r d\bx$, with some parameters in $\omega_{ij}$ treated as the fitting parameters. 
We leave for future work a detailed study with inclusion of ion correlations~\cite{GNE_PRE03, BSK:PRL:2011, XML:PRE:2014}.

\section{Steady States}\label{s:SteadyStates}
\subsection{Steady-State Solutions}\label{ss:ssSolutions}
We find steady-state solutions of the equations~\reff{SPNP1D} by solving equations
\begin{equation}\label{ssSPNP}
\left\{
\begin{aligned}
& \partial_x  \left ( \partial_x c_l + z_l c_l \partial_x \psi+ c_l\sum_{r=1}^Mw_{lr} \partial_x c_r \right) =0 ,~ l=1, \dots, M,\\
&- \partial_{xx} \psi =\kappa \left( \sum_{l=1} ^M z_l c_l + \rho_f \right)\\
\end{aligned}
\right.
\end{equation}
with boundary conditions~\reff{Bcs}. We briefly recall the numerical methods proposed in our previous work~\cite{DSWZhou_CICP18} for solving steady-state solutions.  We first convert the problem into a system of $2M+2$ first order ODEs with the same number of boundary conditions. The ODE system is numerically solved with a Matlab program called BVP4C, in which an ODE system is solved with a collocation method on an adaptive mesh and the resulting nonlinear algebraic equations are iteratively solved by Newton-type methods~\cite{BVP4C}. Given the boundary conditions~\reff{Bcs}, the problem with sign-alternating piecewise constant fixed charges~\reff{rhof} may have multiple steady-state solutions. As the applied voltage $V$ varies, the corresponding current $I$ may become multi-valued, giving rise to an $S$-shaped current-voltage ($I$-$V$) characteristic curve; cf.~Figure~\ref{f:J2Vsize}.  To compute an $I$-$V$ curve, it is helpful to employ the strategy of numerical continuation on applied voltages to get a good initial guess.  However, continuation on $V$ often only finds the low-current branch and high-current branch, and misses the intermediate branch of an $S$-shaped $I$-$V$ curve when multiple solutions are present; cf.~Figure~\ref{f:J2Vsize}. Moreover, the continuation advances with a very small stepsize as $V$ approaches bifurcation points on the curve, where the Jacobian of the discretized nonlinear system becomes close to singular.  To address this issue, the voltage $V$, instead, is viewed as a function of $I$, and the equations~\reff{ssSPNP} are solved with the boundary condition $\psi(1)=V$ in~\reff{Bcs} replaced by a prescribed current $I$.  A complete $S$-shaped $I$-$V$ curve can be obtained by performing continuation on $I$ and collecting the applied voltage by $V=\psi(1)$. See the work~\cite{DSWZhou_CICP18} for more details.

\subsection{Bifurcation Points}
The current-voltage ($I$-$V$) characteristic curve plays a crucial role in the study of ion conductance through ion channel or nanopores.  It is of practical significance to locate bifurcation points on an $S$-shaped $I$-$V$ curve, since the corresponding $V$ values of these critical points are threshold values for hysteresis to take place, and they are also endpoints of intervals for $V$ in which multiple solutions exist. If we assume that an $I$-$V$ curve locally determines a differentiable function of $V$ on $I$, then $\partial_I V =0$ holds at bifurcation points of an $S$-shaped $I$-$V$ curve; cf.~red dots in Figure~\ref{f:J2Vsize}. More generally, it is of interest to locate a point on an $I$-$V$ curve such that the slope, $\partial_I V$,  at the point is given by a known value.
%
%Following previous work \cite{DSWZhou_CICP18} in finding turning points of the steady state PNP, we perform same approach on steady state SPNP equations. It is of primary interest to calculate the the turning points of the $I$-$V$ curve where  $\partial V/\partial I=0$.

At steady states, the current $I$ given in~\reff{Current} is a constant over $\Omega$. Taking derivatives of both sides of the equations~\reff{ssSPNP} with respect to $I$, we have
\begin{equation}\label{1stDEqs}
\left\{
\begin{aligned}
&\partial_x \left( \partial_x \hat{c}_l +z_l \hat{c}_l\partial_x \psi+ z_l c_l \partial_x \hat{\psi} +
\hat{c}_l\sum_{r=1}^Mw_{lr} \partial_x c_r + c_l\sum_{r=1}^Mw_{lr} \partial_x \hat{c}_r
\right)=0,~~ l=1, \cdots, M,\\
& -\partial_{xx} \hat{\psi}=\kappa\sum_{l=1}^M z_l \hat{c}_l,\\
\end{aligned}
\right.
\end{equation}
where $\hat{\psi}:=\partial_I\psi$ and $ \hat{c}_l:= \partial_I c_l$ for $l=1, \dots, M$.
%The boundary conditions are given by
%\begin{equation}\label{1stBcs}
%\left.(\hat{c}_1,\cdots, \hat{c}_M, \hat{\psi})\right|_{x=-1}=(0,\cdots,0,0) ~\mbox{ and }~
%\left.(\hat{c}_1,\cdots, \hat{c}_M, \hat{\psi})\right|_{x=1}=(0,\cdots,0,\partial_I V).
%\end{equation}
To locate a point determined by a given slope value $\partial_I V$, we solve the following boundary-value problem (BVP)
\begin{equation}\label{BVP1st}
\left\{
\begin{aligned}
&\mbox{Equations~\reff{ssSPNP}}, \\
&\mbox{Equations~\reff{1stDEqs}}, \\
&\left.(c_1,\cdots,c_M, \psi, \hat{c}_1,\cdots, \hat{c}_M, \hat{\psi})\right|_{x=-1}=(c_1^L, \cdots,c_M^L, 0, 0,\cdots,0,0), \\
&\left.(c_1,\cdots,c_M,\hat{c}_1,\cdots, \hat{c}_M, \hat{\psi})\right|_{x=1}=(c_1^R, \cdots,c_M^R,0,\cdots,0,\partial_I V).\\
\end{aligned}
\right.
\end{equation}
%consists of the equations~\reff{ssSPNP}, \reff{1stDEqs}, and boundary conditions
%\begin{equation}\label{IVinit}
%\left\{
%\begin{aligned}
%&- \partial_{xx} \psi =  \kappa\left(\sum_{l=1} ^M z_l c_l + \rho_f\right),\\
%&\partial_x\left(\partial_x c_l+z_lc_l\partial_x\psi+ c_l\sum_{r=1}^Mw_{lr} \partial_x c_r\right)=0,~~ l=1, \cdots, M,\\
%& -\partial_{xx} \hat{\psi}=\kappa\sum_{l=1}^M z_l \hat{c}_l,\\
%&\partial_x \left[\partial_x \hat{c}_l +z_l \hat{c}_l\partial_x \psi+ z_l c_l \partial_x \hat{\psi} + \sum_{r=1}^M \left(\omega_{lr}\hat{c}_l\partial_x c_r + \omega_{lr}c_l\partial_x \hat{c}_r \right)\right]=0,~~ l=1, \cdots, M,\\
%\end{aligned}
%\right.
%\end{equation}
%and boundary conditions
%\begin{equation}\label{BoudI}
%\begin{aligned}
%&\left.(c_1,\cdots,c_M, \psi, \hat{c}_1,\cdots, \hat{c}_M, \hat{\psi})\right|_{x=-1}=(c_1^L, \cdots,c_M^L, 0, 0,\cdots,0,0) \\
%&\left.(c_1,\cdots,c_M,\hat{c}_1,\cdots, \hat{c}_M, \hat{\psi})\right|_{x=1}=(c_1^R, \cdots,c_M^R,0,\cdots,0,\partial_I V).\\
%\end{aligned}
%\end{equation}
Note that we do not specify the applied voltage $V$ as a boundary condition in the BVP~\reff{BVP1st}. Instead, the applied voltage can be found after solving the BVP. Since $\sum_{l=1}^M z_l \partial_I J_l =1$, we can convert the BVP into a system of $4M+3$ first order ODEs with the same number of boundary conditions, and numerically solve the ODEs again with the BVP4C program. See our previous work~\cite{DSWZhou_CICP18} for more details.  Of particular interest are the bifurcation points that are determined by $\partial_I V =0$ on an $S$-shaped $I-V$ curve.  With found bifurcation points, we are able to determine threshold applied voltages for hysteresis, as well as intervals for $V$ in which the steric PNP equations have multiple steady states.

Consider an $I$-$V$ curve that locally determines a twice differentiable function of $V$ on $I$. It is desirable to compute the second derivative $\partial_{II} V$ that reveals local shape (concavity) of the curve. In particular, the second derivative enables prediction of approximate local bifurcation diagram close to bifurcation points.  Taking second order derivatives of both sides of the equations~\reff{ssSPNP} with respect to $I$, we have
\begin{equation}\label{2ndDEqs}
\left\{
\begin{aligned}
&\partial_x \left[\partial_x \hat{\hat{c}}_l +2z_l \hat{c}_l\partial_x \hat{\psi} + z_l \hat{\hat{c}}_l\partial_x \psi + z_l c_l \partial_x \hat{\hat{\psi}}  + \sum_{r=1}^M \omega_{lr} \left(2\hat{c}_l\partial_x \hat{c}_r + c_l\partial_x \hat{\hat{c}}_r + \hat{\hat{c}}_l\partial_x c_r  \right) \right]=0, \\
%~~ \mbox{ for } l=1,2,\cdots,M. \\
& -\partial_{xx} \hat{\hat{\psi}}=\kappa\sum_{l=1}^M z_l \hat{\hat{c}}_l,
\end{aligned}
\right.
\end{equation}
%\begin{equation}\label{2ndDEqs}
%\left\{
%\begin{aligned}
%&\partial_x \left[\partial_x \hat{\hat{c}}_l +2z_l \hat{c}_l\partial_x \hat{\psi} + z_l \hat{\hat{c}}_l\partial_x \psi + z_l c_l \partial_x \hat{\hat{\psi}}   \right.\\
%&\left. \qquad\qquad\qquad\qquad + \sum_{r=1}^M \omega_{lr} \left(2\hat{c}_l\partial_x \hat{c}_r + c_l\partial_x \hat{\hat{c}}_r + \hat{\hat{c}}_l\partial_x c_r  \right) \right]=0~~ \mbox{ for } l=1,2,\cdots,M. \\
%& -\partial_{xx} \hat{\hat{\psi}}=\kappa\sum_{l=1}^M z_l \hat{\hat{c}}_l,
%\end{aligned}
%\right.
%\end{equation}
where $\hat{\hat{\psi}}:=\partial_{II}\psi$ and $\hat{\hat{c}}_l := \partial_{II} c_l$ for $l=1, \dots, M$.  Here $c_l$, $\hat{c}_l$, $\psi$, and $\hat{\psi}$ in \reff{2ndDEqs} are obtained by numerically solving the BVP~\reff{BVP1st}.  The boundary conditions are given by
\begin{equation}\label{2ndBcs}
\left.(\hat{\hat{c}}_1,\cdots, \hat{\hat{c}}_M, \hat{\hat{\psi}})\right|_{x=-1}=(0,\cdots,0,0) ~\mbox{ and }~
\left.(\hat{\hat{c}}_1,\cdots, \hat{\hat{c}}_M)\right|_{x=1}=(0,\cdots,0).
\end{equation}
%\begin{equation*}
%\begin{aligned}
%&(\hat{\hat{c}}_1,\cdots, \hat{\hat{c}}_M, \hat{\hat{\psi}})=(0,\cdots,0,0), \mbox{ at } x=-1,\\
%&(\hat{\hat{c}}_1,\cdots, \hat{\hat{c}}_M)=(0,\cdots,0), \mbox{ at } x=1.\\
%\end{aligned}
%\end{equation*}
By the fact that $$\sum_{l=1}^M z_l \partial_{II} J_l =0,$$ we can convert the equations~\reff{2ndDEqs} into a system of $2M+1$ first order ODEs with the same number of boundary conditions, and numerically solve the ODEs with the BVP4C program. After solving, we have $\partial_{II} V = \hat{\hat{\psi}}(1)$. When a bifurcation point is interested, we first solve the BVP~\reff{BVP1st} with $\partial_I V =0$ to get $c_l$, $\hat{c}_l$, $\psi$, and $\hat{\psi}$, and then compute $\partial_{II} V$ to characterize the local bifurcation diagram close to the bifurcation point.

\section{Linear Stability Analysis}\label{s:LinStbAn}
We present a linear stability analysis on steady-state solutions of the steric PNP equations~\reff{SPNP1D}.  We linearize the nonlinear equations~\reff{SPNP1D} around a steady-state solution denoted by $\bar{\psi}(x)$ and $\bar{c}_l(x)$, which is obtained by solving the equations~\reff{ssSPNP} as described in Section~\ref{ss:ssSolutions}. We assume that the solution $(c_1, \cdots, c_M, \psi)$ to the equations~\reff{SPNP1D} is given by
%To determine the linear stability of multiple equilibrium solutions $\bar{\psi}(x), \bar{c}_l(x), 1\leq l\leq M$, we add small perturbations $$(\delta c_1,\cdots,\delta c_M, \delta \psi)=(\delta c_1(x),\cdots,\delta c_M(x), \delta \psi(x))e^{\lambda t}.$$
%The solutions of SPNP equations are found to be
\begin{equation*}
\left(
\begin{aligned}
&c_1(x, t)\\
&~~~~\vdots\\
&c_M(x, t)\\
&\psi(x, t)\\
\end{aligned}
\right)
=
\left(
\begin{aligned}
&\bar{c}_1(x)\\
&~~~~\vdots\\
&\bar{c}_M(x)\\
&\bar{\psi}(x)\\
\end{aligned}
\right)
+e^{\lambda t}
\left(
\begin{aligned}
&\delta c_1(x)\\
&~~~~\vdots\\
&\delta c_M(x)\\
&\delta\psi(x)\\
\end{aligned}
\right) + \cdots,
\end{equation*}
where $|\delta c_l | \ll 1 $ and $|\delta \psi| \ll 1$, and dots represent terms of higher orders as $|\delta c_l |$ and $|\delta \psi |$ go to zero. We also assume that the initial condition is given by
\[
c_l(x,0) = \bar{c}_l(x) + \delta c_l (x) ~~\mbox{and} ~~ \psi (x,0) = \bar{\psi}(x) + \delta \psi (x).
\]
By the Dirichlet boundary conditions~\reff{Bcs}, we have homogeneous boundary conditions for $\delta c_l $ and $\delta \psi$:
\begin{equation}\label{HDBCs}
\begin{aligned}
%(\bar{c}_1, \bar{c}_2,\cdots,\bar{c}_M, \bar{\psi})&=(c_{1}^L, c_{2}^L, \cdots,c_{M}^L,0)~~\mbox{at}~ x=-1,\\
%(\bar{c}_1, \bar{c}_2,\cdots,\bar{c}_M, \bar{\psi})&=(c_{1}^R, c_{2}^R, \cdots,c_{M}^R,V)~~\mbox{at}~ x=1,\\
\left. (\delta c_1 , \delta c_2,\cdots, \delta c_M, \delta \psi)\right|_{x=\pm1}&=(0, 0,\cdots,0, 0).\\
\end{aligned}
\end{equation}
To the leading order, we recover the equations~\reff{ssSPNP} for the steady state. Comparing the first-order terms, we obtain a linearized system
\begin{equation*}\label{linearized}
\left\{
\begin{aligned}
&\lambda \delta c_l= \gamma_l \partial_x\left[\partial_{x} \delta c_l+z_l\bar{c}_l\partial_x \delta \psi+z_l\delta c_l\partial_x \bar{\psi}+\sum_{r=1}^M w_{lr}\left(\delta c_l\partial_x \bar{c}_r+\bar{c}_l\partial_x \delta c_r\right)\right],~~1\leq l\leq M,\\
&-\partial_{xx} \delta \psi=\kappa \sum_{l=1}^M z_l \delta c_l.
\end{aligned}
\right.
\end{equation*}
This system can be expressed in a matrix form
\begin{equation}\label{eigenfunction}
A\left(
\begin{aligned}
&\delta c_1 \\
&\delta c_2\\
&~~\vdots\\
&\delta c_M
\end{aligned}
\right)=\lambda\left(
\begin{aligned}
&\delta c_1 \\
&\delta c_2\\
&~~\vdots\\
&\delta c_M\\
\end{aligned}
\right),
\end{equation}
where
\begin{equation}\label{OperatorMatrix}
A=\left(
\begin{array}{cccc}
A_{11} & A_{12} & \cdots & A_{1M}\\
A_{21} & A_{22} & \cdots & A_{2M}\\
\vdots & \vdots & \vdots & \vdots\\
A_{M1} & A_{M2} & \cdots & A_{MM}\\
\end{array}
\right)
\end{equation}
with operators
\begin{equation*}
\begin{aligned}
&\frac{A_{ii}}{\gamma_i}=\partial_{xx}-z_i^2\kappa\partial_x \bar{c}_i\partial_x(\partial_{xx})^{-1}-z_i^2\kappa \bar{c}_i+z_i\partial_x\bar{\psi}\partial_x+z_i\partial_{xx} \bar{\psi}+w_{ii}\partial_x \bar{c}_i \partial_x +w_{ii}\bar{c}_i\partial_{xx}\\
&\qquad \qquad \qquad+\sum_{r=1}^M w_{ir} \left(\partial_x \bar{c}_r \partial_x +\partial_{xx}\bar{c}_r \right)~~\mbox{for}~ i= 1, 2, \cdots, M,\\
&\frac{A_{ij}}{\gamma_i}=-z_iz_j\kappa\partial_x\bar{c}_i\partial_x(\partial_{xx})^{-1}-z_iz_j\kappa\bar{c}_i+w_{ij}\partial_x \bar{c}_i \partial_x +w_{ij}\bar{c}_i\partial_{xx}~~\mbox{ for}~i\neq j.\\
\end{aligned}
\end{equation*}

To numerically compute $\lambda$, we develop a finite difference scheme to discretize the operator matrix \reff{OperatorMatrix}. We introduce a uniform mesh with grid points given by
$$x_i=-1+i h,\qquad i=0,1,\cdots, N+1,$$ where grid spacing $h=\frac{2}{N+1}.$ We denote by $\delta c_{l,i}$ and $\delta \psi_{i}$ the numerical approximation of $\delta c_l(x_i)$ and $\delta \psi(x_i)$ for $ i=0,1,\cdots, N+1$, respectively. By the boundary conditions~\reff{HDBCs}, we have $\delta c_{l,0}=\delta c_{l,N+1}=0$ and $\delta \psi_{0}= \delta \psi_{N+1} = 0$.  We take interior grid points $\left\{x_i\right\}_{i=1}^N$ as our computational mesh, and define
\begin{equation}\label{InteriorGrids}
\begin{aligned}
&\widehat{\delta c_l}=(\delta c_{l,1},\delta c_{l,2},\cdots,\delta c_{l,N})^{T},~~~1\leq l\leq M,\\
&\widehat{\delta \psi}=(\delta \psi_{1},\delta \psi_{2},\cdots,\delta \psi_{N})^{T}.
\end{aligned}
\end{equation}
The first derivative is approximated by a second-order central differencing scheme, which corresponds to a differentiation matrix
\begin{equation*}%\label{OperatorMatrix}
D_{N+2}=\frac{1}{2h}\left(
\begin{array}{ccccc}
0 & 2 &      &  & 0\\
-1 & 0 & 1  &  &\\
     &\ddots & \ddots & \ddots & \\
     &          & -1 & 0 &  1 \\
0    &    &      & -2  & 0
\end{array}
\right)_{(N+2)\times(N+2).}
\end{equation*}
The second derivative corresponds to a differentiation matrix $D^2_{N+2}$.
%For eigenvalue problem \reff{eigenfunction} with homogeneous Dirichlet boundary conditions
%$$(\delta c_1,\delta c_2,\cdots,\delta c_M, \delta \psi)=(0,0,\cdots,0) \mbox{ at } x=-1, 1.$$
%Here the subscript denotes the ion species, we take interior grid points $x_2, x_3,\cdots, x_{N}$ as computational grids,
%Here the superscript denotes the ion species and subscript denotes spatial grid.
% What's more, the value on the boundary grids $\delta c^l_1, \delta c^l_{N+1}, (1\leq l \leq M), \delta \psi_1, \delta \psi_{N+1}$ are zero.
To approximate the eigenvalue problem \reff{eigenfunction} on interior grid points, we only make use of $\tilde{D}_N$ and $\tilde{D}^2_N$ obtained by stripping the first and last rows and columns of $D_{N+2}$ and $D^2_{N+2}$, i.e., in Matlab notations,
$$\tilde{D}_N=D_{N+2}(2:N+1, 2:N+1)~\mbox{~and~} \tilde{D}^2_N=D^2_{N+2}(2:N+1, 2:N+1).$$
The operators $\partial_{x}$ and $\partial_{xx}$ with homogenous Dirichlet boundary conditions can be approximated by $\tilde{D}_N$ and $\tilde{D}^2_N$, respectively. Similarly, the operator $(\partial_{xx})^{-1}$ is approximated by $(\tilde{D}^2_N)^{-1}$.  Consequently, the $A_{ij}$ blocks can be approximated as
\begin{equation*}\label{operatormatrix}
\begin{aligned}
&\frac{\mathcal{A}_{ii}}{\gamma_i}\approx\tilde{D}^2_N-z_i^2\kappa\tilde{D}_N \bar{c}_i\tilde{D}_N(\tilde{D}^2_N)^{-1}-z_i^2\kappa \bar{c}_i+z_i\tilde{D}_N\bar{\psi}\tilde{D}_N+z_i\tilde{D}^2_N \bar{\psi}\\
&\qquad \qquad \qquad+w_{ii}\tilde{D}_N \bar{c}_i \tilde{D}_N +w_{ii}\bar{c}_i\tilde{D}^2_N+\sum_{r=1}^M w_{ir}(\tilde{D}_N \bar{c}_r \tilde{D}_N +\tilde{D}^2_N\bar{c}_r ),~~~1\leq i\leq M,\\
&\frac{\mathcal{A}_{ij}}{\gamma_i}\approx-z_iz_j\kappa\tilde{D}_N\bar{c}_i\tilde{D}_N(\tilde{D}^2_N)^{-1}-z_iz_j\kappa\bar{c}_i+w_{ij}\tilde{D}_N \bar{c}_i \tilde{D}_N +w_{ij}\bar{c}_i\tilde{D}^2_N,~~~i\neq j.\\
\end{aligned}
\end{equation*}

We assemble the vectors $\widehat{\delta c_l}$ defined in~\reff{InteriorGrids} for $M$ species of ions, and define
$${\mathbf {\delta c}} := \sum_{l=1}^M \be_l \otimes \widehat{\delta c_l},$$
where $\be_l$ is a unit column vector whose the $l$-th element is the only nonzero element being one, and $\otimes$ represents a Kronecker product. We recast the matrix $A$ in~\reff{eigenfunction} as a matrix $\mathbf {A}: \R^{MN} \to \R^{MN}$ given by
$$\mathbf {A}=\sum_{i,j=1}^M \left( \be_i \otimes  \be_j^T \right) \otimes \mathcal{A}_{ij}.$$
Therefore, the eigenvalue problem~\reff{eigenfunction} can be approximated by
\begin{equation}\label{DisEigen}
\mathbf {A} {\mathbf {\delta c}} =\lambda {\mathbf {\delta c}}.
\end{equation}
Such an eigenvalue problem determines the linear stability of the steric PNP equations~\reff{SPNP1D} around a steady state as follows: $\mbox{Re}(\lambda)>0$ describes unstable modes, and $\mbox{Re}(\lambda)<0$ describes stable modes. For short time, the coupling of perturbations of infinitesimal amplitude, as well as their impact on steady-state solutions, is still negligible. However, if $\mbox{Re}(\lambda)>0$ for one certain mode, perturbations associated with the mode grow exponentially and quickly reach an amplitude where nonlinearity has to be taken into account. In such a nonlinear regime, the above linear stability analysis is no longer applicable. To study the nonlinear dynamics, e.g., transition from one steady state to another, we resort to numerical simulations of the steric PNP equations~\reff{SPNP1D} starting from a steady state.

%Typically, as can beseen in Figure\reff{f:J2Vsize}, there is an unstable branch joining two turning points, which connects to stable branches above and below at the turning points.
%\subsection{Time-dependent SPNP equations}
%To verify that the solutions deviate from the middle unstable branch to the stable branch, we solve the time dependent SPNP equations with BVP4C, implicitly.
To avoid possible instability arising from temporal discretization, we propose a stable, implicit time integration scheme for the steric PNP equations~\reff{SPNP1D}:
\begin{equation}\label{PNPtime}
\left\{
\begin{aligned}
&\frac{c_l^{n+1}-c_l^n}{\Delta t}=\gamma_l \partial_x \left( \partial_x   c_l^{n+1} + z_lc_l^{n+1} \partial_x  \psi^{n+1} + c_l^{n+1}\sum_{r=1}^M\omega_{lr} \partial_x c_r^{n+1} \right),\\
&-\partial_{xx} \psi^{n+1}=\kappa\left(\sum_{l=1}^M z_l c_l^{n+1}+\rho_f\right),
\end{aligned}
\right.
\end{equation}
with boundary conditions~\reff{Bcs}.  Here we denote by $c_l^n(x)$ and $\psi^n(x)$ semi-discrete approximations of $c_l(t_n, x)$ and $\psi(t_n, x)$, where $t_n=n\Delta t$ and $\Delta t$ is a uniform time step size.  Such a semi-discrete scheme can be converted into an ODE system and solved with the BVP4C, as described in Section~\ref{ss:ssSolutions}.

\section{Numerical Results}\label{s:NumRes}
In our numerical simulations, we numerically solve the equations~\reff{ssSPNP} with certain boundary conditions to find an $I$-$V$ curve. To understand more on an $S$-shaped $I$-$V$ curve, we solve the BVP~\reff{BVP1st} with $\partial_{I} V =0$ to locate bifurcation points, and solve the equations~\reff{2ndDEqs} at bifurcations points  with boundary conditions~\reff{2ndBcs} to find $\partial_{II} V$, which can be further used to predict local bifurcation diagram close to the bifurcations points. Moreover, we solve the eigenvalue problem~\reff{DisEigen} to study linear stability of steady states. In nonlinear regime, we numerically investigate transition dynamics from unstable steady states to stable ones with the time integration scheme~\reff{PNPtime}.

Unless otherwise stated, we consider a charged system that consists of  binary monovalent electrolytes, i.e., $M=2$, $z_1=1$, and $z_2=-1$, and fixed charges
\[
\rho_f(x)=
\left\{
\begin{aligned}
&0,  \qquad  & x \in [-1, -0.2), \\
&-50,               & x \in [-0.2, 0.4), \\
&20,               & x \in [0.4, 0.8), \\
&-50,               & x \in [0.8, 0.95), \\
&0,               & x \in [0.95, 1). \\
\end{aligned}
\right.
\]
We consider a steric interaction matrix $W$ with diagonal elements $w_{11}=w_{22}=0.1$ and subdiagonal elements $w_{12}=w_{21}=0.01$. We take $\gamma_1=\gamma_2=1$, $\kappa$=100, and boundary concentrations $c_1^L=1$,  $c_2^L=1$, $c_1^R=0.5$, and $c_2^R=0.5$.
%$\rho_1^f=0$, $\rho_2^f=-50$, $\rho_3^f =20$, $\rho_4^f=-50$, $\rho_5^f=0$,  and $L_1=0.8$, $L_2=0.6$, $L_3=0.4$, $L_4=0.15$, $L_5=0.05$, $w_{11}=0.1$, $w_{22}=0.1$, $w_{12}=0.01$, $w_{21}=0.01$. and interaction matrix

%, where $\alpha$ and $\beta$ shall be prescribed later.

\subsection{Hysteresis and Bistability}\label{s:IVcurves}
\begin{figure}[htbp]
    \centering
	\includegraphics[scale=0.52]{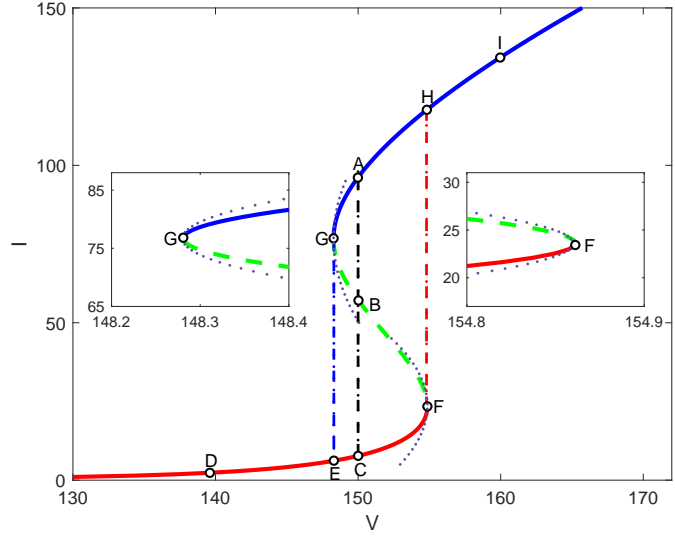}
\caption{An $S$-shaped $I$-$V$ curve obtained by solving the equations~\reff{ssSPNP}. Linear stability analysis reveals that the steady-state solutions on the upper (blue solid) and lower (red solid) branches are stable, and those on the intermediate (green dashed)  branches are unstable. The second derivative is given by $\partial_{II} V=0.005$ at the upper bifurcation point ($G$) that is located at $(148.28, 76.69)$. The second derivative is given by $\partial_{II} V=-0.011$ at the lower bifurcation point ($F$) that is located at $(154.86, 23.61)$. There are three steady-state solutions, labeled by $A$, $B$, and $C$, when the applied voltage $V=150$. The zoomed-in inset presents local bifurcation diagram, in which the black dotted parabolas are predicted by the second derivatives at bifurcation points. }
\label{f:J2Vsize}
\end{figure}
We study the current-voltage relation for voltage-induced hysteresis of ionic conductance. i.e., the $I$-$V$ curve as shown in Figure~\ref{f:J2Vsize}. We first numerically solve the equations~\reff{ssSPNP} with continuation on applied voltages. When the applied voltage increases gradually,  the current ascends along a low-current branch  (from $D$ to $C$ on the solid red curve). As the voltage approaches the critical value corresponding to $F$, the iterations for solving the equations slow down due to a more and more singular Jacobian matrix, and continuation has to advance with small increment. After the applied voltage exceeds a critical value about $V=154.86$, the current suddenly jumps from the low-current branch at the bifurcation point $F$ to a high-current branch at $H$, switching the ionic conductance state. In the other round of sweeping with decreasing applied voltages from high values,  the current follows a totally different path (from $I$ to $A$ on the solid blue curve) and suddenly switches at the bifurcation point $G$ with $V=148.28$ to $E$ on the low-current branch, exhibiting a typical hysteresis loop. This demonstrates that ion conductance depends on the applied voltage as well as its history values, showing a memory effect.  We remark that this is reminiscent of the voltage-induced gating mechanism~\cite{GavishLiuEisenberg_JPCB18}, though our theory is purely deterministic. To explore more of the intermediate region of two branches, we view the applied voltage as a function of the current and solve the equations~\reff{ssSPNP} with a boundary condition prescribing the current, instead of applied voltages. Such treatment of the boundary condition stablizes the numerical simulation of steady-state solutions that are shown to be unstable.  As shown by the green dashed curve in Figure~\ref{f:J2Vsize}, an intermediate branch connecting the low-current and high-current branches can be found by continuation on the current.  The complete $S$-shaped curve features hysteretic response of ion current to varying applied voltages.

\begin{figure}[htbp]
    \centering
	\includegraphics[scale=0.52]{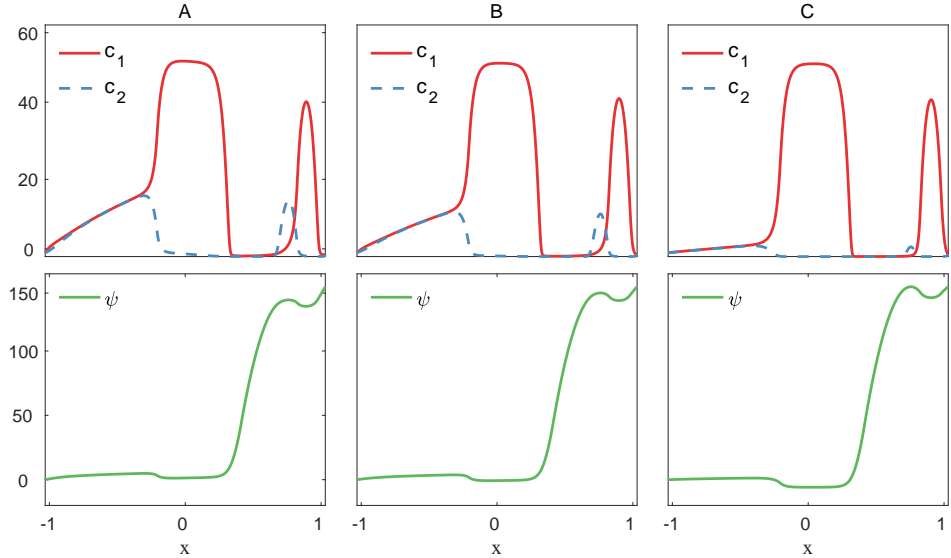}
    \caption{The solution profiles of the electrostatic potential and ionic concentrations, corresponding to $A$, $B$, and $C$ in the Figure~\ref{f:J2Vsize} when $V = 150$. }
\label{f:3Solutions}
\end{figure}
For applied voltages in the interval $(148.28, 154.86)$, there exist three steady-state solutions of different ion conductance levels. For instance, three steady-state solutions, labeled by $A$, $B$, and $C$, are found for $V=150.$ The panel presented in Figure~\ref{f:3Solutions} displays profiles of the electrostatic potential and ionic concentrations corresponding to these three points.  We observe that the solutions have large variations at the locations where the fixed charges are discontinous. The ions distribute mainly according to the fixed charges due to electrostatic interactions. Comparing with the low-current state, both cations and anions have larger concentrations along the channel in the high-current conductance state.

\begin{figure}[htbp]
\centering
\includegraphics[scale=.62]{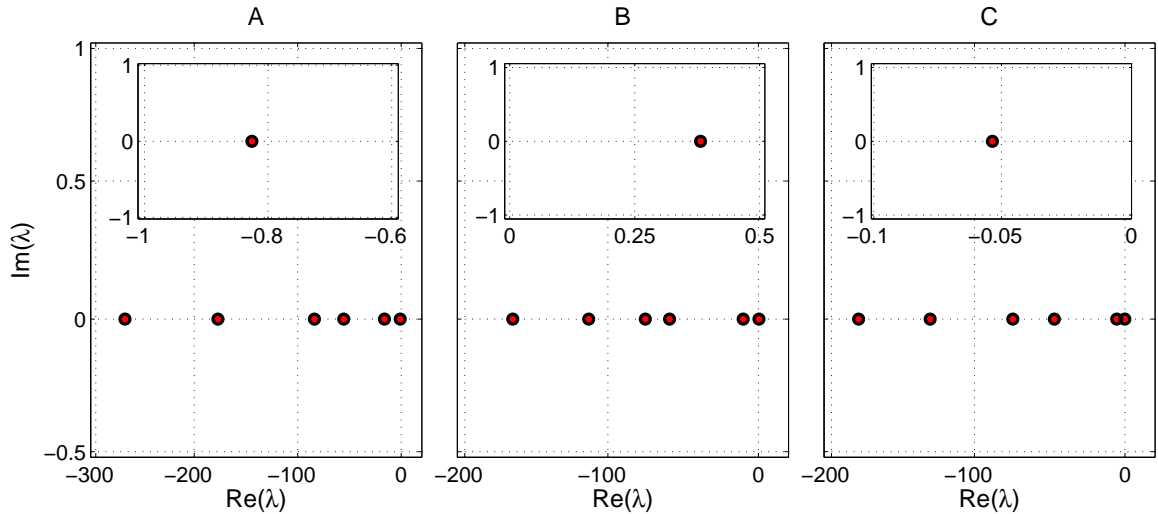}
\caption{The eigenvalues with the largest real parts of the eigenvalue problem~\reff{DisEigen} at steady-state solutions corresponding to the $A$, $B$, and $C$ in the Figure~\ref{f:J2Vsize}. Insets are zoomed-in plots of the eigenvalue with the largest real part. }
\label{f:eigenvalues}
\end{figure}
%The dominant eigenvalues for the perturbed system from stability analysis of points A, B, C from figure \reff{f:J2Vsize} are plotted in the figure \reff{f:eigenvalues}. Panels A and C feature negative eigenvalues, showing that points A and C in Figure\reff{f:J2Vsize} left panel are on a stable branch. Panel B of Figure\reff{f:eigenvalues} features positive dominant eigenvalue, indicating that point B in Figure\reff{f:J2Vsize} left panel is on an unstable branch.

We perform linear stability analysis on the steady-state solutions on the lower, intermediate, and upper branches of the $S$-shaped $I$-$V$ curve by solving the eigenvalue problem~\reff{DisEigen}. The sign of the computed eigenvalue with the largest real part, i.e., the dominant eigenvalue, for the perturbed system determines the linear stability. We present in Figure~\ref{f:eigenvalues} six eigenvalues with the largest real parts of the eigenvalue problem~\reff{DisEigen} at steady-state solutions, corresponding to the $A$, $B$, and $C$ in the Figure~\ref{f:J2Vsize}. We find that the real part of the dominant eigenvalue is negative for both $A$ and $C$, and is positive for $B$. This reveals that the steady-state solutions at $A$ and $C$ are linearly stable, but that at $B$ is linearly unstable. Analogously, we consider the real part of the dominant eigenvalue for other points on the $I$-$V$ curve. Remarkably, we find that the steady-state solutions on both the lower and upper branches shown in blue solid curves are stable, but those on the intermediate branch shown in red dashed curves are unstable, exhibiting a classical bistable diagram. This agrees with our expectation of bistability as seen in typical hysteresis loops. Also, the instability explains why it is hard to capture the steady-state solutions on the intermediate branch in our numerical computations~\cite{DSWZhou_CICP18}.

The stable and unstable branches connect at bifurcation points marked by red solid dots, which are located by solving the BVP~\reff{BVP1st} with $\partial_{I} V =0$. Further numerical calculations based on the equations~\reff{2ndDEqs} at bifurcations points with boundary conditions~\reff{2ndBcs} find the second derivative of $V$ with respect to $I$:  $\partial_{II} V=-0.011$ at the lower bifurcation point and $\partial_{II} V=0.005$ at the upper bifurcation point. Such second derivative information helps predict local bifurcation diagram close to the bifurcation points. The dotted parabolas shown in the inset of Figure~\ref{f:J2Vsize} demontrate that the predicted parabola agrees well locally with the one calculated by using steady-state solutions.

%As we can see in the left figure \reff{f:J2Vsize}, the upper and lower branch (blue, solid lines) are stable and the middle branch (red, dashed line) is unstable. Solving the equations \reff{IVinit} by BVP4C, we can get two turning points labeled by red dots, at $(V,I)$ coordinates $(148.28,76.69)$ and $(154.86,23.61)$. The embedded figures clearly exhibit zoomed-in features of curve near turning points.  Solutions to the equations \reff{IVAug} yields approximation of the shape around turning points by parabolas given by $\delta^2 V=0.005 \delta I^2$ and $\delta^2 V=-0.011 \delta I^2$ respectively.
%
%The six panels on the right of figure \reff{f:J2Vsize} display profiles of electrostatic potential and concentration corresponding to points A, B, C in the left figure \reff{f:J2Vsize} when $V = 150$.  We observe that the solutions of electrostatic potential resemble each other with the same potential differences, $V = 150$. The concentrations $c_1$ and $c_2$ for each case all satisfy the positive constraint.

%In this example, we study the stability of current through an ion channel with two ionic species to varying applied voltages.\cite{HungMihn_arXiv15} and \cite{Gavish_PhysD18} have investigated the stability of stationary state of SPNP equations.

\subsection{Convergence Test}
\begin{table}[htbp]
\centering
\begin{tabular}{ccccccc}
\hline  \hline
 \multirow{2}{*}{$h$} &\multicolumn{2}{c}{A} &\multicolumn{2}{c}{B}&\multicolumn{2}{c}{C} \\
 \cline{2-3}  \cline{4-5}  \cline{6-7}
 & Error in $\lambda_{\rm Max}^A$ & Order & Error in $\lambda_{\rm Max}^B$ & Order & Error in $\lambda_{\rm Max}^C$ & Order\\
 \hline
 %$\frac{1}{650}$ & 3.14$\times 10^{-2}$ & -& 1.93$\times 10^{-1}$ & -& 3.91$\times 10^{-3}$ & - \\
 %$\frac{1}{700}$ & 2.76$\times 10^{-1}$ & 1.78& 1.66$\times 10^{-1}$ & 2.05& 3.34$\times 10^{-1}$ & 2.11 \\
 $\frac{1}{750}$ & 2.42$\times 10^{-2}$ & -& 1.44$\times 10^{-1}$ & 2.07& 2.88$\times 10^{-1}$ & -\\
 $\frac{1}{800}$ & 2.15$\times 10^{-2}$ & 1.88& 1.25$\times 10^{-1}$ & 2.09 & 2.51$\times 10^{-1}$ & 2.15\\
 $\frac{1}{850}$ & 1.91$\times 10^{-2}$ & 1.96& 1.11$\times 10^{-1}$ & 2.11 & 2.20$\times 10^{-1}$ & 2.17\\
 $\frac{1}{900}$ & 1.70$\times 10^{-2}$ & 1.97& 9.79$\times 10^{-2}$ & 2.14& 1.94$\times 10^{-1}$ & 2.19\\
 \hline  \hline
\end{tabular}
\caption{The error and convergence order of $\lambda_{\rm Max}$, which is the eigenvalue with the largest real part, for the eigenvalue problem at $A$, $B$, and $C$ in the Figure~\ref{f:J2Vsize}.}\label{t:convergence}
\end{table}
We perform convergence tests with various mesh resolutions to further validate the proposed numerical approaches on  linear stability analysis of the steric PNP equations around steady-state solutions. Table~\ref{t:convergence} lists the error and convergence order of $\lambda_{\rm Max}$, which is the eigenvalue with the largest real part, for the eigenvalue problem at $A$, $B$, and $C$ in the Figure~\ref{f:J2Vsize}. Table~\ref{t:Vectconvergence} lists $l^\infty$ errors and convergence order of the eigenvector, corresponding to the eigenvalue $\lambda_{\rm Max}$, for the ionic concentrations and electrostatic potential at $A$, $B,$ and $C$.  The error is obtained by comparing against a reference solution that is computed on a highly refined mesh with $h=\frac{1}{3000}$. Since second-order central differencing is employed to discretize the differential operators in~\reff{OperatorMatrix}, we anticipate a second-order convergence rate of our numerical approaches.  From the tables, we can see that the numerical error decreases robustly as the mesh refines, and the numerical convergence order is indeed roughly about two for both eigenvalues and eigenvectors.
\begin{table}[htbp]
\centering
\begin{tabular}{cccccccc}
 \hline \hline
 & $h$ & $l^\infty$-error in $\delta c_1$ & Order & $l^\infty$-error in $\delta c_2$ & Order & $l^\infty$-error in $\delta \psi$ & Order\\
 \hline
 \multirow{4}{*}{$A$}
 %$\frac{1}{650}$ & 1.50$\times 10^{-3}$ & -& 4.63$\times 10^{-4}$ & -& 3.05$\times 10^{-4}$ & - \\
 %$\frac{1}{700}$ & 1.30$\times 10^{-3}$ & 2.11& 3.96$\times 10^{-4}$ & 2.12& 2.61$\times 10^{-4}$ & 2.14 \\
 &$\frac{1}{750}$ & 1.10$\times 10^{-3}$ & -& 3.41$\times 10^{-4}$ & -& 2.25$\times 10^{-4}$ & -\\
 &$\frac{1}{800}$ & 9.42$\times 10^{-4}$ & 2.14& 2.97$\times 10^{-4}$ & 2.15 & 1.95$\times 10^{-4}$ & 2.17\\
 &$\frac{1}{850}$ & 8.27$\times 10^{-4}$ & 2.15& 2.61$\times 10^{-4}$ & 2.16 & 1.71$\times 10^{-4}$ & 2.17\\
 &$\frac{1}{900}$ & 7.30$\times 10^{-4}$ & 2.19& 2.30$\times 10^{-4}$ & 2.19& 1.51$\times 10^{-4}$ & 2.21\\
% $\frac{1}{950}$ & 6.48$\times 10^{-4}$ & 2.20& 2.04$\times 10^{-4}$ & 2.20& 1.34$\times 10^{-4}$ & 2.21 \\
 %$\frac{1}{1000}$ & 5.77$\times 10^{-4}$ & 2.24& 1.82$\times 10^{-4}$ & 2.24& 1.19$\times 10^{-4}$ & 2.25 \\
 %\hline  \hline
 %\multirow{2}{*}{$h$} &\multicolumn{6}{c}{B}  \\
% \cline{2-7}
%    & error in $c_1$ & order & error in $c_2$ & order & error in $\psi$ & order\\
 \hline
  \multirow{4}{*}{$B$}
 %$\frac{1}{650}$ & 8.31$\times 10^{-4}$ & -& 4.20$\times 10^{-4}$ & -& 9.01$\times 10^{-5}$ & - \\
 %$\frac{1}{700}$ & 7.12$\times 10^{-4}$ & 2.11& 3.61$\times 10^{-4}$ & 2.12& 7.70$\times 10^{-5}$ & 2.12 \\
 &$\frac{1}{750}$ & 6.16$\times 10^{-4}$ & -& 3.13$\times 10^{-4}$ & -& 6.64$\times 10^{-5}$ & -\\
 &$\frac{1}{800}$ & 5.37$\times 10^{-4}$ & 2.12& 2.74$\times 10^{-4}$ & 2.08 & 5.78$\times 10^{-5}$ & 2.15\\
 &$\frac{1}{850}$ & 4.72$\times 10^{-4}$ & 2.13& 2.41$\times 10^{-4}$ & 2.10 & 5.06$\times 10^{-5}$ & 2.17\\
 &$\frac{1}{900}$ & 4.17$\times 10^{-4}$ & 2.16& 2.13$\times 10^{-4}$ & 2.13& 4.47$\times 10^{-5}$ & 2.19\\
% $\frac{1}{950}$ & 3.70$\times 10^{-4}$ & 2.18& 1.90$\times 10^{-4}$ & 2.16 & 3.96$\times 10^{-5}$ & 2.22 \\
% $\frac{1}{1000}$ & 3.31$\times 10^{-4}$ & 2.21& 1.70$\times 10^{-4}$ & 2.18& 3.53$\times 10^{-1}$ & 2.24 \\
% \hline  \hline
% \multirow{2}{*}{$h$} &\multicolumn{6}{c}{C}  \\
% \cline{2-7}
%    & error in $c_1$ & order & error in $c_2$ & order & error in $\psi$ & order\\
 \hline
  \multirow{4}{*}{$C$}
 %$\frac{1}{650}$ & 2.20$\times 10^{-3}$ & -& 2.60$\times 10^{-3}$ & -& 4.13$\times 10^{-4}$ & - \\
 %$\frac{1}{700}$ & 1.90$\times 10^{-3}$ & 2.14& 2.20$\times 10^{-3}$ & 2.18& 3.49$\times 10^{-4}$ & 2.28 \\
 &$\frac{1}{750}$ & 1.60$\times 10^{-3}$ &  -      & 1.90$\times 10^{-3}$ & -& 2.98$\times 10^{-4}$ & -\\
 &$\frac{1}{800}$ & 1.40$\times 10^{-3}$ & 2.17& 1.60$\times 10^{-3}$ & 2.20 & 2.57$\times 10^{-4}$ & 2.27\\
 &$\frac{1}{850}$ & 1.20$\times 10^{-3}$ & 2.19& 1.40$\times 10^{-3}$ & 2.21 & 2.24$\times 10^{-4}$ & 2.28\\
 &$\frac{1}{900}$ & 1.10$\times 10^{-3}$ & 2.21& 1.30$\times 10^{-3}$ & 2.23& 1.97$\times 10^{-4}$ & 2.29\\
% $\frac{1}{950}$ & 9.73$\times 10^{-4}$ & 2.23& 1.10$\times 10^{-4}$ & 2.25& 1.74$\times 10^{-4}$ & 2.30 \\
% $\frac{1}{1000}$ & 8.67$\times 10^{-4}$ & 2.25& 9.85$\times 10^{-4}$ & 2.27& 1.54$\times 10^{-4}$ & 2.31 \\
 \hline \hline
\end{tabular}
\caption{The $l^\infty$ error and convergence order of the eigenvector corresponding to $\lambda_{\rm Max}$ at the $A$, $B$, and $C$ in the Figure~\ref{f:J2Vsize}.}\label{t:Vectconvergence}
\end{table}

%To verify the accuracy, we numerically solve the problem by using various spatial step size $h$. Here the mesh ratio is chosen for the purpose of accuracy test.
%Table \reff{t:Vectconvergence} lists $l^\infty$ errors and convergence order of eigenvector corresponding to maximum eigenvalue for ionic concentration and electrostatic potential at  $A, B,$ and $C$. We observe that error decreases as the mesh refines, and that convergence order of eigenvectors corresponding to maximum eigenvalue for ionic concentration and electrostatic potential are both about 2.

\subsection{Nonlinear Transition Dynamics}
The linear stability analysis is no longer applicable, when long-time dynamics of the steric PNP equations are concerned. To understand how the unstable steady-state solution evolves to stable ones, we numerically investigate the nonlinear transition dynamics with the time integration scheme~\reff{PNPtime}. The numerical simulations start from a steady-state  solution corresponding to $B$ on the unstable branch of the $S$-shaped $I$-$V$ curve shown in Figure~\ref{f:J2Vsize}. Since numerical noise provides broadband perturbations, noise associated to the mode of the eigenvalue with $\text{Re}(\lambda_{\rm Max}^B)>0$ will grow exponentially for small time.
\begin{figure}[htbp]
\centering
\includegraphics[scale=.53]{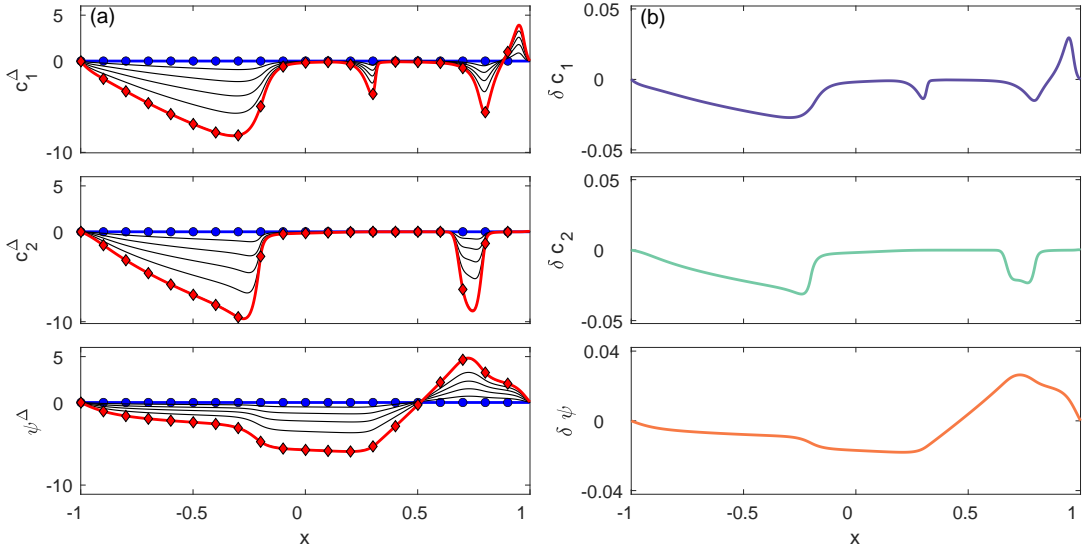}
\caption{(a) Snapshots of the evolution of $\psi^{\Delta}(t, x)$ and $c_l^{\Delta}(t, x)$ ($l=1,\cdots,M$) starting from the unstable steady-state solution at $B$ (blue lines marked with dots) to a final state corresponding to $C$ (red curves marked with diamonds). (b): A normalized eigenvector corresponding to the eigenvalue with the largest real part at $B$.}
\label{f:EorrBC}
\end{figure}% B to C
To examine the transition starting from an unstable solution, we define variations of $\psi$ and $c_l$ as
$$\psi^{\Delta}(t, x):=\psi(t, x)-\psi^B(x),~ c_l^{\Delta}(t, x):=c_l(t, x)-c_l^B(x),~~l=1,\cdots,M, $$
where $\psi(t, x)$ and $c_l(t, x)$ are numerical solutions to~\reff{PNPtime} with initial conditions $\psi^B(x)$ and $c_l^B(x)$ that are a steady-state solution corresponding to $B$ in Figure~\ref{f:J2Vsize}. As shown in Figure~\ref{f:EorrBC} (Left), both $\psi^{\Delta}(t, x)$ and $c_l^{\Delta}(t, x)$ start from zero (blue lines marked with dots), and evolve to final states (red curves marked with diamonds), corresponding to a stable solution at $C$ in Figure~\ref{f:J2Vsize}. In addition, Figure~\ref{f:EorrBC} (Right) shows a normalized eigenvector corresponding to the eigenvalue with the largest real part at $B$. Clearly, we find that the evolution of concentrations and the electrostatic potential follows exactly in the direction dictated by the eigenvector, especially in the early stage. This is consistent with our linear stability analysis that the perturbations associated to the mode of the eigenvalue with the largest real part lead transition dynamics.

\begin{figure}[htbp]
\centering
\includegraphics[scale=.53]{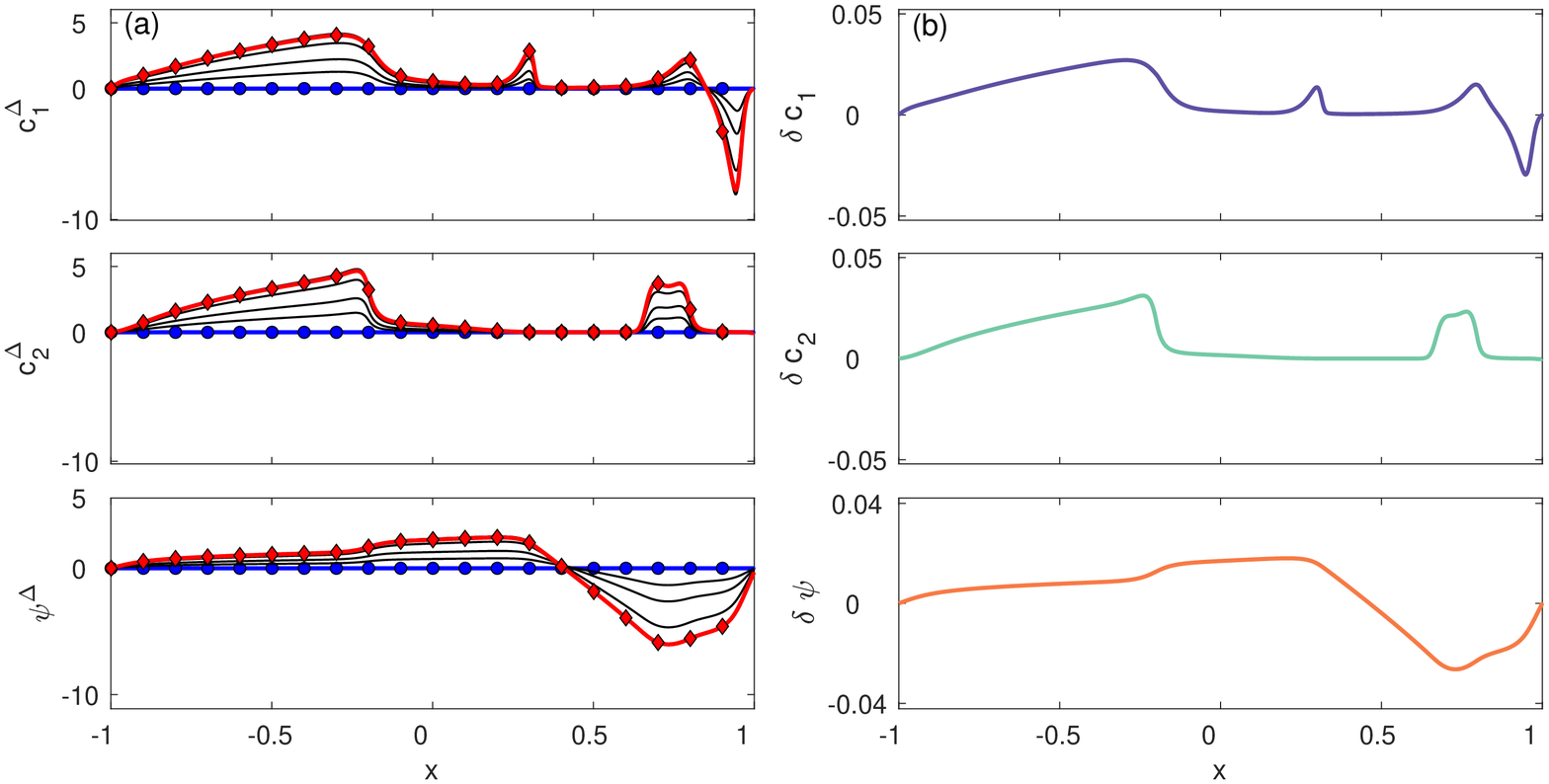}
\caption{(a): Snapshots of the evolution of $\psi^{\Delta}(t, x)$ and $c_l^{\Delta}(t, x)$ ($l=1,\cdots,M$) starting from the initial condition~\reff{PertB}  (blue lines marked with dots)  to a final state corresponding to $A$ (red curves marked with diamonds). (b): Another normalized eigenvector corresponding to the eigenvalue with the largest real part at $B$. }
\label{f:EorrBA}
\end{figure}% B to A

To study transition dynamics from $B$ to $A$, numerical integration starts from an initial condition
\begin{equation}\label{PertB}
\psi(0, x)=\epsilon \psi^A(x) + (1-\epsilon) \psi^B(x),~ c_l(0, x) = \epsilon  c_l^A(x) + (1-\epsilon) c_l^B(x) ,~~l=1,\cdots,M,
\end{equation}
where $\epsilon = 0.001$. Such an initial condition is obtained by perturbing the steady-state solution at $B$ slightly toward that at $A$. Figure~\ref{f:EorrBA} (Left) shows the evolution of $\psi^{\Delta}(t, x)$ and $c_l^{\Delta}(t, x)$ from the initial condition~\reff{PertB} to a final state at $A$.   Figure~\ref{f:EorrBA} (Right) shows another normalized eigenvector corresponding to the eigenvalue with the largest real part at $B$. Notice that this normalized eigenvector points in an opposite direction contrast to the one showed in Figure~\ref{f:EorrBC} (Right). Again, we can observe that the evolution of concentrations and the electrostatic potential accords with the given eigenvector, agreeing with our linear stability analysis.

%From figure \reff{f:Dynamics}, we can find the orientation of the solutions of unstable point to the solutions of stable point match with the eigenvector corresponding to maximum eigenvalue of the unstable point. Snapshots are starting from $(c^1_{\rm diff}, c^2_{\rm diff}, \psi_{\rm diff})=(0, 0, 0)$ initial conditions(blue dots). What's more, we can get
%\begin{equation}
%\begin{aligned}
%\psi_{\rm diff}&\propto\delta\psi e^{\lambda T},\\
%c^l_{\rm diff}&\propto\delta c_l e^{\lambda T} \mbox{ for } l=1,2.
%\end{aligned}
%\end{equation}
%
%After a short period of time, the solutions of perturbed system \reff{PNPtime} resemble the dominant eigenvector of linearized system \reff{eigenfunction}. In the final state the solutions have relaxed to the stable branch equilibrium solutions with the same boundary conditions as the initial, unstable equilibrium.

\subsection{Concentration-Induced Hysteresis}
\begin{figure}[htbp]
    \centering
    \includegraphics[scale=.6]{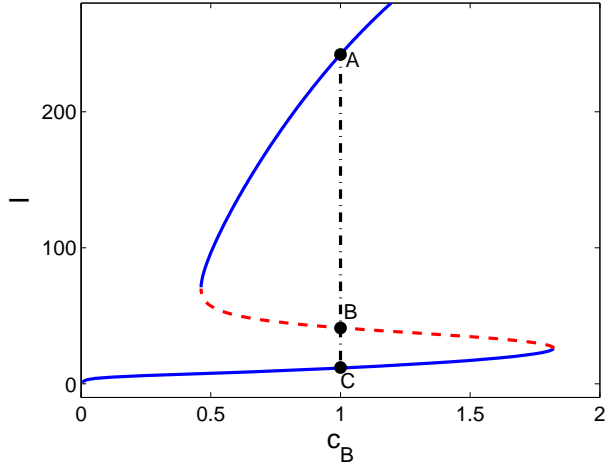}
    \caption{A current-concentration curve showing concentration-induced hysteresis. The steady-states on the blue solid curve are linearly stable, and those on the red dashed curve are linearly unstable.}
    \label{f:2ionsIC}
\end{figure}
In this example, we investigate the effect of boundary concentrations on the existence and stability of multiple steady-state solutions to the steric PNP equations. We set boundary concentrations $c_l(-1)=c_l^L$ for $l=1, 2$ at the left end where the electrostatic potential is grounded, and set the relation $c_1(1)=c_2(1)$ for boundary concentrations at the right end where the potential $V = 150$. We define the variable $c_B:=c_1(1)$. Similar to the $S$-shaped $I$-$V$ curve shown in Figure~\ref{f:J2Vsize}, two rounds of sweeping with an increasing and decreasing $c_B$ are used to capture low-current and high-current branches in the current-concentration curve shown in Figure~\ref{f:2ionsIC}. To find the intermediate branch, we prescribe the current $I$ as a boundary condition and treat $c_B$ as a variable to be determined.  See our previous work~\cite{DSWZhou_CICP18} for more details.  Such a strategy helps complete the whole current-concentration curve with a hysteresis loop, which indicates that the channel conductance state not only depends on intracellular and extracellular ionic concentrations but also their history values.

\begin{figure}[htb]
    \centering
    \includegraphics[scale=.65]{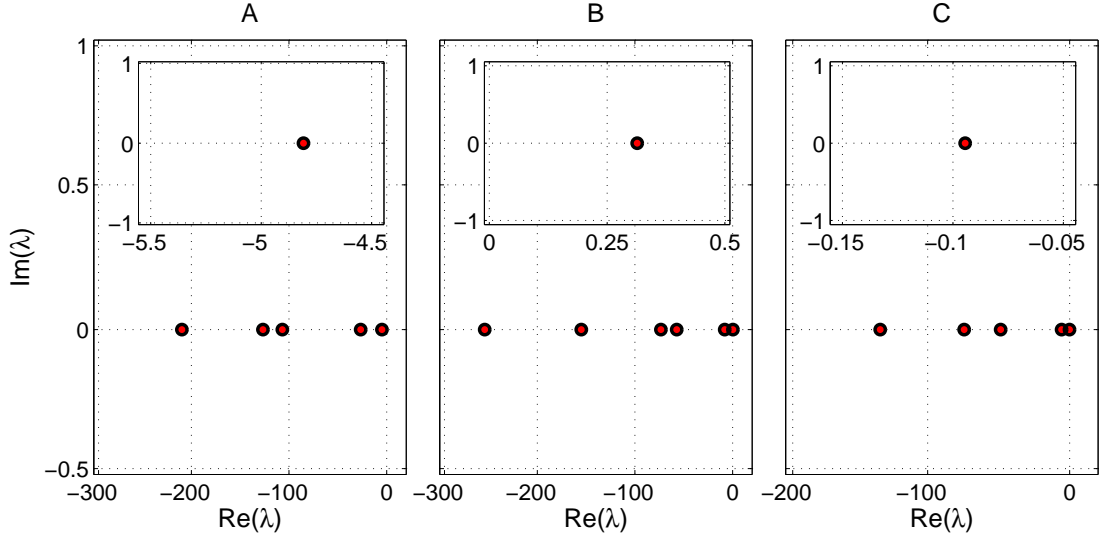}
    \caption{The eigenvalues with the largest real parts of the eigenvalue problem~\reff{DisEigen} at steady-state solutions corresponding to the points $A$, $B$, and $C$ in the current-concentration curve shown in Figure~\ref{f:2ionsIC}. Insets are zoomed-in plots of the eigenvalue with the largest real part. }\label{f:ICeigenvalues}
\end{figure}

In addition, we study linear stability of steady state solutions on the current-concentration curve by solving the eigenvalue problem~\reff{DisEigen}. For the points $A$, $B,$ and $C$ on the curve with $c_B=1$, we can see from Figure~\ref{f:ICeigenvalues} that the dominant eigenvalue has a negative real part for $A$ and $C$, and a positive real part for $B$. This indicates that the steady-state solutions at $A$ and $C$ are linearly stable and that at $B$ is linearly unstable. Similarly, we check the real part of the dominant eigenvalue for the whole current-concentration curve. Results demonstrate that the steady-state solutions on blue solid curves are linearly stable, and that on the dashed intermediate branch in red are unstable, showing a typical bistable hysteresis loop.

\section{Conclusions}\label{s:Conclusions}
In this work, we have studied linear stability of multiple steady-state solutions to the steric PNP equations.  The current-voltage ($I$-$V$) characteristic curve has been determined by solving steady states to the steric PNP equations with Dirichlet boundary conditions arising from ion channels. We have numerically found multiple steady-state solutions that lead to an $S$-shaped current-voltage and current-concentration curves showing hysteresis loops. The hysteretic response of ion conductance to applied voltages and boundary concentrations demonstrates that ion conductance state depends on their currently applied values as well as  their history values, i.e, the memory effect.  We have proposed boundary value problems to locate of bifurcation points and to predict local bifurcation diagram near bifurcation points on the $S$-shaped $I$-$V$ curve.  To understand the stability of steady-state solutions that are only numerically available, we have performed linear stability analysis that is based on finite difference discretization of differential operators.  The linear stability analysis has unveiled that the $S$-shaped $I$-$V$ curve has two linearly stable branches with a high-conductance state and low-conductance state, and has one linearly unstable intermediate branch. This presents a classical hysteresis diagram with bistable behavior. Furthermore, extensive numerical tests have shown that the numerical method proposed in the linear stability analysis is  second-order accurate. In addition, we have numerically studied the nonlinear transition dynamics from unstable steady-state solutions to stable solutions. It has been found that the transition follows the direction predicted by eigenvectors associated to the dominant eigenvalue in the linear stability analysis. The numerical approaches developed in this work are quite general and useful, in the sense that they can be applied to study linear stability of other time-dependent problems around their steady-state solutions that are only numerically available.

\bigskip
\noindent{\bf Acknowledgments.}
J. Ding was supported by National Natural Science Foundation of China (No.~21773165, 11771318, and 11790274) and Postgraduate Research~\& Practice Innovation Program of Jiangsu Province.  H. Sun was supported by Simons Foundation Collaborative Grant (No.~522790).  S. Zhou was supported by the grants NSFC 21773165, Natural Science Foundation of Jiangsu Province, Young Elite Scientist Sponsorship Program by Jiangsu Association for Science and Technology, and National Key R\&D Program of China  (No. 2018YFB0204404).

%%\bibliographystyle{plain}
%\bibliographystyle{unsrt}
%%\bibliographystyle{alpha}
%\bibliography{PNPStb}

}
\end{document}